\documentclass{moriond}

\usepackage{amssymb}

\def\be{\begin{equation}}
\def\ee{\end{equation}}
\def\bea{\begin{eqnarray}}
\def\eea{\end{eqnarray}}

\newcommand{\Photo}{}

\begin{document}
{\normalsize\rightline{DESY-23-076}} 
\vspace*{4cm}
\title{
Discovery potential for axions in Hamburg}

\author{ A. Ringwald }

\address{Deutsches Elektronen-Synchrotron  DESY, Notkestr. 85, 22607 Hamburg, Germany}

\maketitle\abstracts{
We review the motivation for axions, discuss benchmark axion models, and 
report on the ongoing and planned axion experiments in Hamburg and their discovery potential. 
}

\section{Motivation}

The Standard Model (SM) of particle physics is extraordinarily successful. However, it 
lacks a particle candidate for dark matter. Moreover, it does not explain why the $\bar\theta$-angle, 
which measures the strength of CP violation in strong interactions and determines e.g. the 
electric dipole moment of the neutron (nEDM), $d_n\sim 10^{-16}\, \bar\theta\,e\, {\rm cm}$ , is so tiny,
$|\bar\theta|<10^{-10}$, as inferred from the experimental upper bound on the latter. 

Intriguingly, an extension of the SM by a global axial symmetry -- the Peccei-Quinn (PQ) symmetry~\cite{Peccei:1977hh} -- 
which is spontaneously broken at a scale $v_{\rm PQ}$  
solves both these puzzles in one go. The field corresponding to the pseudo Nambu-Goldstone boson arising from 
PQ symmetry breaking -- the axion~\cite{Weinberg:1977ma,Wilczek:1977pj} ($a$) -- acts as a space-time dependent $\bar \theta$-angle, 
$\theta_a (x)=a(x)/f_a$, that is it 
interacts in the low energy effective Lagrangian with the gluonic field strengths $G_{\mu\nu}^b$ and 
their duals $\tilde{G}^{b,\mu\nu}$ as  
${\mathcal L}\supset \theta_a \,\frac{\alpha_s}{8\pi}  G_{\mu\nu}^b \tilde{G}^{b,\mu\nu}$,
where $\alpha_s=g_s^2/(4\pi)$ is the strong coupling and $f_a\propto v_{\rm PQ}$ the axion decay constant. Non-perturbative QCD dynamics leads then to an effective potential $V(\theta_a )$ which has a minimum 
at vanishing field value. Correspondingly, the vacuum expectation value of the axion vanishes, $\langle\theta_a\rangle =0$, implying 
a vanishing nEDM, $d_n\propto \langle \theta_a\rangle =0$, and thus solving the strong CP puzzle. 
Moreover, for a large enough decay constant, $f_a\gtrsim 10^8$\,GeV, the axion is produced in the early universe very efficiently by 
non-thermal mechanisms, such as vacuum realignment~\cite{Preskill:1982cy,Abbott:1982af,Dine:1982ah} 
and the decay of topological defects, and may well be the main constituent of the cold dark matter observed in the 
universe today.

\section{Benchmark Axion Models}

\begin{figure}[h]
\centerline{\includegraphics[width=0.4\linewidth]{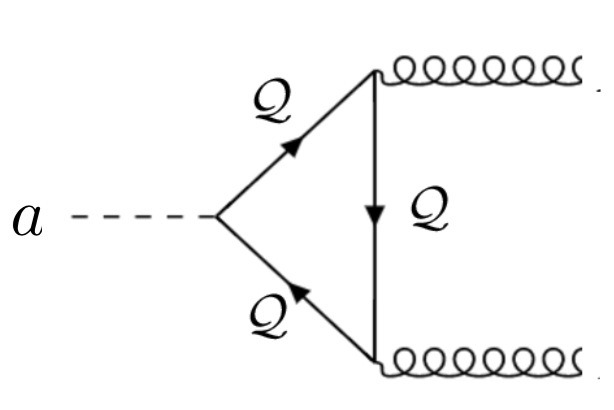}}
\caption[]{Triangle loop diagram giving rise to the axion coupling to gauge fields.}
\label{fig:electromagnetic_coupling_axion}
\end{figure}

A particular simple field-theoretic realisation of the PQ mechanism has been constructed by Kim~\cite{Kim:1979if},
and Shifman, Vainshtein and Zakharov~\cite{Shifman:1979if} 
(KSVZ). It consists of a SM-singlet complex scalar field $\sigma$, featuring a global $U(1)_{\rm PQ}$ symmetry, 
$\sigma \to e^{i\alpha} \sigma$, which is spontaneously broken in the vacuum, and an exotic quark ${\mathcal Q} = {\mathcal Q}_L + {\mathcal Q}_R$, on which the PQ symmetry acts in an axial manner,  
${\mathcal Q}_L \to e^{i\alpha/2} {\mathcal Q}_L$,  ${\mathcal Q}_R \to e^{-i\alpha/2} {\mathcal Q}_R$, with most general 
Lagrangian
\begin{equation}
\mathcal{L}_{\rm KSVZ} = |\partial_\mu \sigma|^2 
- \lambda_{\sigma} \left( \left| \sigma \right|^{2} -\frac{v_{\rm PQ}^2}{2} \right)^{\!\! 2} 
+ \overline {\mathcal Q}\, i \gamma^\mu  
D_\mu 
{\mathcal Q} 
- \left( y_{\mathcal Q} \overline {\mathcal Q}_L {\mathcal Q}_R \sigma + {h.c.} \right),
\end{equation}
with covariant derivative $D_\mu = \partial_\mu -ig_s T^a G_\mu^a -i q_{\mathcal Q} e A_\mu$, 
where $G_\mu^a$ are the gluonic gauge fields, $T^a$ are the colour gauge group generators, 
$q_{\mathcal Q}$ a possible electric charge of the exotic quark and $A_\mu$ the photonic gauge field. 
Decomposing the complex scalar in terms of polar coordinates, 
$\sigma (x) = \frac{1}{\sqrt{2}} \left( v_\sigma + \rho (x)\right) {\rm e}^{i a(x)/v_{\rm PQ}}$, 
it is easily seen that this model features three particles beyond the SM: two massive ones -- 
the excitation of the modulus field $\rho$ with mass $m_\rho =\sqrt{2\lambda_\sigma} v_{\rm PQ}$ and 
the exotic quark with mass $m_{\mathcal Q} =\frac{y_{\mathcal Q}}{\sqrt{2}} v_{\rm PQ}$ -- 
and a massless one -- the excitation of the angular field $a$. 
For large PQ breaking scale, $v_{\rm PQ}\gg v \simeq 246\,{\rm GeV}$, the former two new particles have masses far above the electroweak scale and may be integrated out, if we are interested in an effective description of the physics at energies below the PQ breaking scale.   
This results in the following low energy effective Lagrangian for the angular field $a$, 
\begin{equation}
\mathcal{L}_{\rm KSVZ} \simeq 
\frac{1}{2} \partial_\mu a \partial^\mu a  
+  a\,  \frac{2 N}{v_{\rm PQ}} \, \frac{g_s^2}{32\pi^2} \,
G \tilde G
+  a\,  \frac{2 E}{v_{\rm PQ}} \, \frac{e^2}{32\pi^2} \,
F \tilde F \, ,
\end{equation}
where $F$ is the electromagnetic (EM) field strength tensor and $\tilde F$ its dual. 
Here, the coupling to the gluonic fields arises from the triangle loop diagram in 
Fig.~\ref{fig:electromagnetic_coupling_axion}, where the curly lines are to be interpreted 
%
%
as gluonic gauge fields, while the coupling to the photonic fields arises from the 
same diagram if the curly lines are interpreted as photonic gauge fields. 
The coefficient $N$ is determined by the difference of 
PQ charges of the left and right handed components of the exotic quark, 
$N= (X_{\mathcal{Q}_L}-X_{\mathcal{Q}_R})/2$, which 
results in $N= ({1}/{2}-(-{1}/{2}))/2=1/2$.
So, in fact, $\theta_a=a/f_a$, with $f_a=v_{\rm PQ}/(2 N)=v_{\rm PQ}$, acts as a 
space-time dependent $\bar\theta$-angle and the strong CP puzzle is solved: the angular 
field in the KSVZ model corresponds to the axion.
The coefficient $E$ is obtained as $E=3  (X_{\mathcal{Q}_L}-X_{\mathcal{Q}_R}) q_{\mathcal Q}^2=3q_{\mathcal Q}^2$. 

%
\begin{figure}[t]
\centerline{\includegraphics[width=\linewidth]{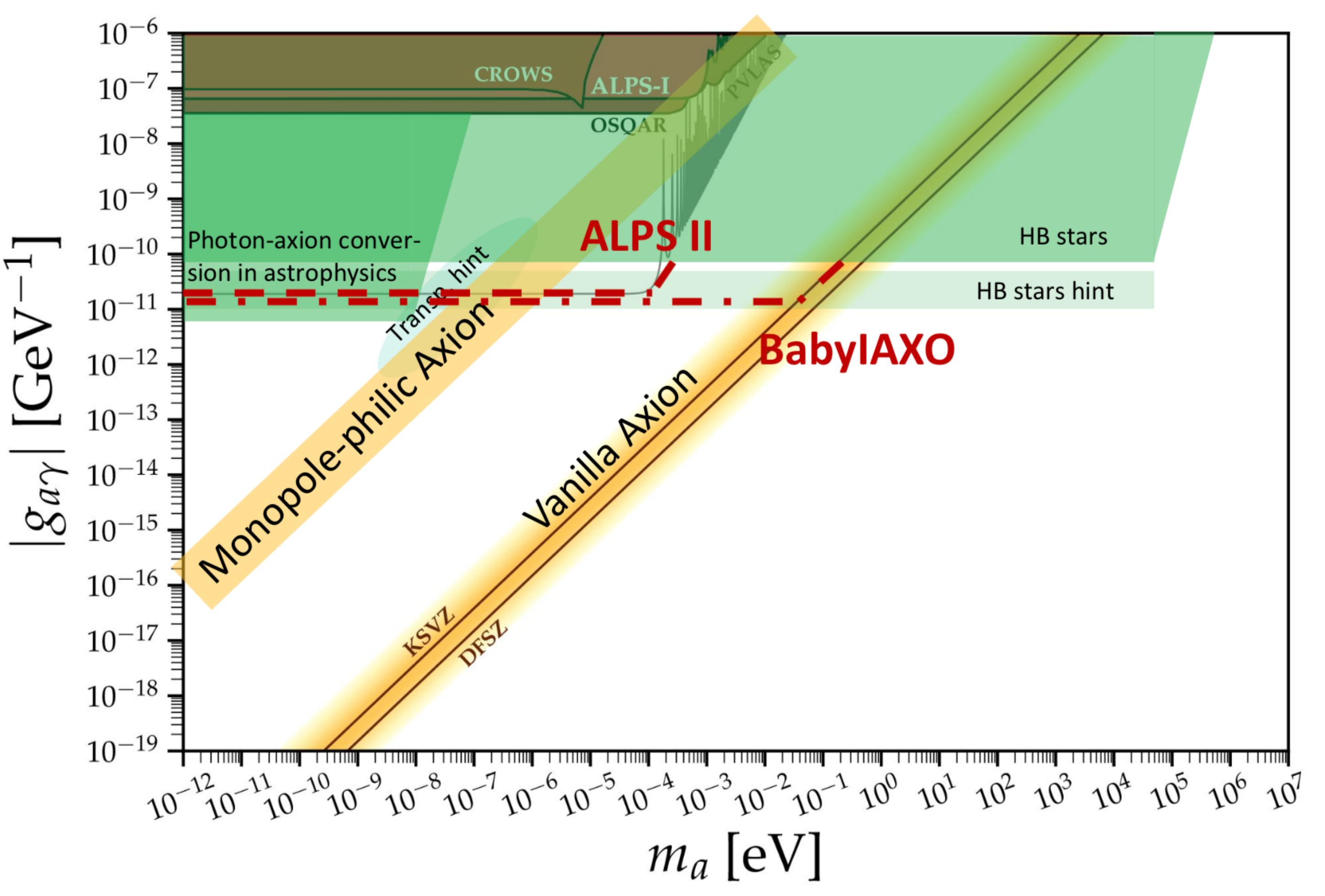}}
\caption[]{EM couplings of the axion versus its mass. Ranges of axion model predictions are displayed as 
yellow bands, excluded regions by astrophysical considerations (energy loss of Horizontal Branch (HB) stars, photon-axion conversion in
astrophysics) in darker green, regions hinted at by astrophysical anomalies (anomalous transparency of universe for $\gamma$ rays, too much energy loss of HB stars) in light green. Excluded regions by past LSW experiments are shown as red filled regions, the projected sensitivity of ALPS II and BabyIAXO as red dashed and dashed-dottet lines, respectively.}
\label{fig:electromagnetic_couplings_vs_mass}
\end{figure}
%

In axion experiments we are interested in physics below the QCD scale. After integrating out the  
gluon fields one obtains 
\begin{equation}
\mathcal{L}_{\rm KSVZ} \simeq
 \frac{1}{2}\, \partial_\mu a\, \partial^\mu  
-\frac{1}{2}  m_{a}^2 a^2  
+ 
\frac{1}{4} g_{a\gamma\gamma}\, a   
\, F_{\mu\nu} \tilde{F}^{\mu\nu}  \,.
\end{equation}
The axion mass $m_a$ and its EM coupling $g_{a\gamma\gamma}$ are given by 
\begin{equation}
\label{eq:mass_coupling}
m_a   \simeq \frac{\sqrt{z}}{1+z}\,\frac{m_\pi \, f_\pi}{f_a } 
\approx 6\ {\rm meV} \left( \frac{10^9\,{\rm GeV}}{f_a}\right), \hspace{3ex} 
g_{a\gamma\gamma}= \frac{\alpha}{2\pi f_a} \left(\frac{E}{N}-\frac{2}{3}\frac{4+z}{1+z}\right)
\,,
\end{equation}
respectively, 
with $z=m_u/m_d\approx 1/2$, in terms of the up (down) quark mass $m_{u(d)}$ and $\alpha=e^2/(4\pi)$.
The axion mass and the second term in $g_{a\gamma\gamma}$ arise from the mixing of the axion with the neutral pion. 
The line in Fig.~\ref{fig:electromagnetic_couplings_vs_mass} labeled as `KSVZ' displays the prediction of 
$g_{a\gamma\gamma}$, as a function of $m_a$, from Eq.~\ref{eq:mass_coupling}, for $q_{\mathcal Q}=0$. 
Varying $q_{\mathcal Q}$ gives rise to the yellow `band' of predictions~\cite{DiLuzio:2016sbl} labeled as `Vanilla Axion' in Fig.~\ref{fig:electromagnetic_couplings_vs_mass}. 
But this does not exhaust all the possible variants of the KSVZ model. In fact, the exotic quark may carry also 
a magnetic charge $g_{\mathcal Q}$. In this case the KSVZ axion model would solve not only the strong CP puzzle, but also the 
charge quantisation puzzle~\cite{Dirac:1931kp,Schwinger:1966nj,Zwanziger:1970hk}. The triangle loop in
Fig.~\ref{fig:electromagnetic_coupling_axion} induces then an EM  coupling~\cite{Sokolov:2021ydn,Sokolov:2021eaz,Sokolov:2022fvs,Sokolov:2023pos}
$g_{aMM} 
\simeq   
 ({\alpha_m}/{2\pi f_a }) 
 ({M}/{N})$, where 
$\alpha_{m} \equiv {g_0^2}/({4\pi})$ and 
$M = 3 g_{\mathcal Q}^2$,
which is, due to charge quantisation,  $e g_0=6\pi n$, $n\in\mathbb{Z}$, parametrically enhanced in comparison to the coupling induced by an electrically charged exotic quark:
${g_{aMM}}/{g_{a\gamma\gamma}} = (\alpha_m/\alpha )(M/E) = (9/4) \, \alpha^{-2}  (g_{\mathcal Q}/q_{\mathcal Q})^2 \sim 10^5$.
The corresponding prediction is shown as a band labeled `Monopole-philic Axion' in Fig.~\ref{fig:electromagnetic_couplings_vs_mass}. 
It overlaps with the regions labeled as `Transparency hint' and `HB hint' in the same figure, demonstrating that such a monopole-philic axion, 
for masses around $\sim 10^{-7}$\,eV, may at the same time also explain the anomalous transparency of the universe for TeV gamma rays~\cite{Meyer:2013pny}  and the anomalous energy losses of horizontal branch stars in globular clusters~\cite{Ayala:2014pea}. 
Moreover, in the pre-inflationary PQ symmetry breaking scenario, an axion in this mass range would very naturally account for 100\% of dark matter~\cite{Borsanyi:2016ksw}.

The axion model constructed by Zhitnitsky~\cite{Zhitnitsky:1980tq} and by Dine, Fischler and Srednicki~\cite{Dine:1981rt} -- the DFSZ 
model -- exploits an extended electroweak Higgs sector -- a two Higgs doublet model -- but no extension of the fermionic sector by an exotic quark. The right-handed fermions, the two Higgs doublets and the complex PQ scalar carry PQ charges. Summing over all PQ charged SM fermions in a triangle loop similar to 
Fig.~\ref{fig:electromagnetic_coupling_axion} leads then to $N_{\rm DFSZ}=3$ and $(E/N)_{\rm DFSZ}=8/3$. 
The line in Fig.~\ref{fig:electromagnetic_couplings_vs_mass} labeled as `DFSZ' displays the corresponding prediction of 
$g_{a\gamma\gamma}$, as a function of $m_a$. 
Unlike the KSVZ model, the DFSZ model has also a tree level coupling of the axion to SM fermions, 
$\mathcal{L}_{\rm DFSZ} \supset ({1}/{2}) \, C_{af}({\partial_\mu a}/{f_{a}})\,\overline{\psi}_f \gamma^\mu\gamma_5 \psi_f$, 
with $C_{au}  = {\cos^2 \beta}/{3}$ and $C_{ad} = C_{ae} =  {\sin^2 \beta}/{3}$, where $\tan\beta = v_u/v_d$ is the ratio of the vacuum expectation values of the two Higgses, with $v=(v_u^2+v_d^2)^{1/2}\simeq 246$\,GeV.

\section{Axion Experiments in Hamburg}

The axion experiments in Hamburg are all exploiting the coupling of the axion to electromagnetism. 
We will describe the search techniques used by these experiments and their discovery potential in the following subsections.

\subsection{Searching for Home-made Axions}

\begin{figure}[h]
\begin{minipage}{0.45\linewidth}
\centerline{\includegraphics[width=0.9\linewidth]{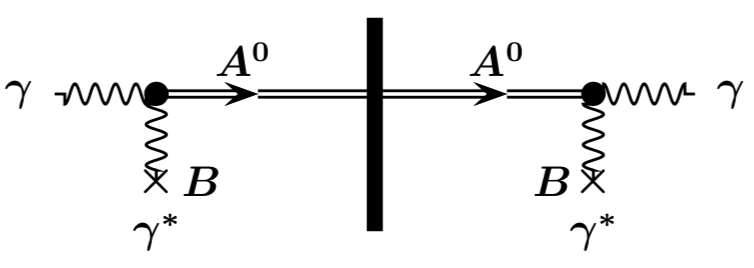}}
\end{minipage}
\hfill
\begin{minipage}{0.55\linewidth}
\centerline{\includegraphics[width=0.9\linewidth]{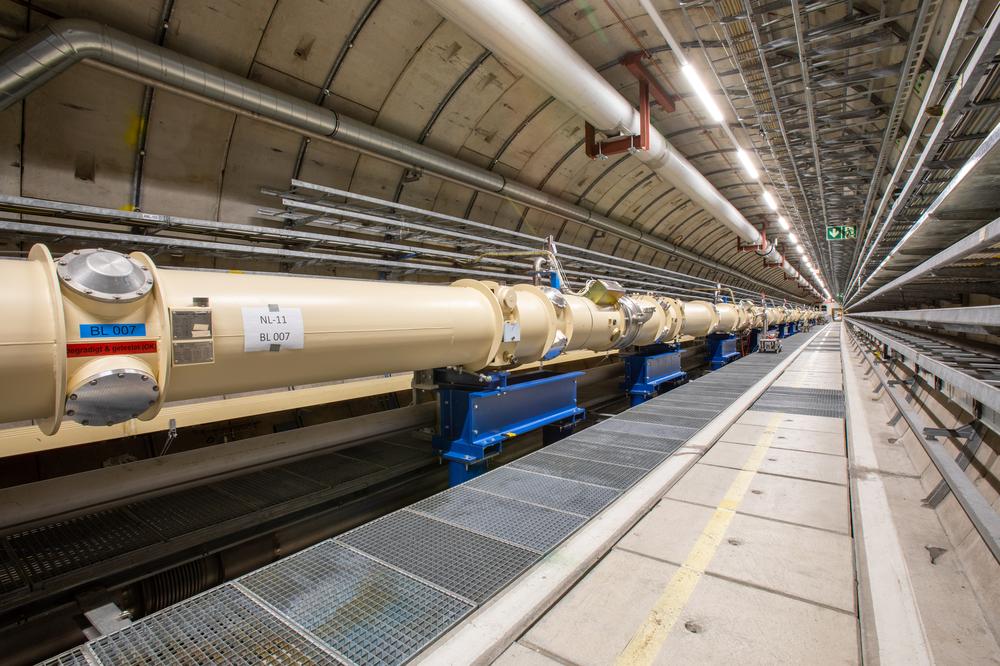}}
\end{minipage}
\caption[]{LSW concept~\cite{Ringwald:2003nsa} ({\em left}) and magnet string of the ALPS experiment in the HERA tunnel 
({\em right}).}
\label{fig:LSW_ALPS_II}
\end{figure}

Experiments exploiting the so-called ``Light Shining through a Wall" (LSW) technique~\cite{Anselm:1985obz,VanBibber:1987rq,Ringwald:2003nsa,Redondo:2010dp,Spector:2023nap} 
are powerful tools to search for axions via 
their coupling $g_{a\gamma\gamma}$ or $g_{aMM}$ in a pure laboratory and thus astrophysical-model independent setup. 
They are based on the fact that an EM field wave sent along a transverse magnetic field may 
convert partially into an axion field wave, and vice versa (Sikivie effect~\cite{Sikivie:1983ip}).
Correspondingly, putting a light-tight wall in the middle of the transverse magnetic
field region, the existence of axions is signaled by an EM wave emerging behind the wall from 
axion - photon conversion, cf. Fig.~\ref{fig:LSW_ALPS_II} (left). 
For light axions, $m_a\ll (4\omega/L_B)^2$, 
where $\omega$ is the photon energy and $L_B$ the length of the magnetic conversion region, the probability 
that a photon converts into an axion is given by 
$P(\gamma \rightarrow a ) \simeq \frac{1}{16} \left( g_{a\gamma\gamma} B L_B\right)^4\simeq P(a \rightarrow \gamma )$, 
where $g_{a\gamma\gamma}$ can be replaced by $g_{aMM}$ for the monopole-philic axion~\cite{Sokolov:2022fvs}. 
The currently best pure laboratory limits on the EM coupling of light axions ($m_a\lesssim 10^{-4}$\, eV) have been achieved 
by the LSW experiments ALPS~\cite{Ehret:2010mh} (``Any Light Particle Search'') at DESY and 
OSQAR~\cite{OSQAR:2015qdv} (``Optical Search of QED vacuum magnetic birefringence, Axion and photon Regeneration") at CERN,
see Fig.~\ref{fig:electromagnetic_couplings_vs_mass}. However, these 
limits are deep in the parameter region excluded by the limits from the non-observation of  beyond the SM energy losses of HB stars, see
Fig.~\ref{fig:electromagnetic_couplings_vs_mass}. 

The LSW experiment ALPS II  in Hamburg (see Fig.~\ref{fig:LSW_ALPS_II} (right)) has been designed~\cite{Bahre:2013ywa} to surpass the latter limit in a model-independent way and thus to dig into previously 
uncharted territory in axion parameter space, checking also the axion explanation of the previously mentioned astrophysical anomalies (see Fig.~\ref{fig:electromagnetic_couplings_vs_mass}). 
The required improvement in sensitivity by a factor of a thousand in comparison to ALPS rests mainly on the ideas to use {\em i)} a string of 
recycled superconducting HERA dipoles in one of the straight sections of the HERA tunnel~\cite{Ringwald:2003nsa} instead of only one such magnet as in ALPS and {\em ii)} an optical cavity also on the after-wall side where the photons are regenerated (Regeneration Cavity (RC)) 
to enhance resonantly their number~\cite{Hoogeveen:1990vq,Fukuda:1996kwa,Sikivie:2007qm} instead of using only one optical cavity before the wall (Production Cavity (PC)) as in ALPS. 

In concreteness, ALPS~II exploits two strings constituted by straightened HERA dipole magnets~\cite{Albrecht:2020ntd} -- 12 before and 12 after the wall -- inside of which the PC and RC, respectively, are located~\cite{Ortiz:2020tgs}. 
The expected power of the regenerated light, for $m_a\lesssim 0.1$\,meV, is given by 
\begin{equation}
    {\mathcal P}_\gamma 
\simeq {\mathcal P}_{\rm PC}  
    \frac{1}{16} (g_{a\gamma \gamma} B L_B)^4  
\beta_{\rm RC}  
\simeq  
    6 \times 10^{-24 } \, {\rm W} \frac{{\mathcal P}_{\rm PC}}{150\,{\rm kW}} 
\frac{\beta_{\rm RC}}{4\times 10^4} 
    \left(\frac{g_{a\gamma \gamma}}{2\times 10^{-11} \rm{~GeV^{-1}}} \frac{B}{5.3 \rm{~T}} \frac{L_B}{105.6 \rm{~m}}\right)^4
\,,
\end{equation}
where ${\mathcal P}_{\rm PC}$ is the total circulating power in the PC and $\beta_{\rm RC}$
the resonant enhancement factor in the RC, implying an expected rate 
$\dot{N}_\gamma =  {\mathcal P}_\gamma/\omega \simeq 3/{\rm day}$ of photons with a wavelength 
$\lambda = 1064$\,nm, corresponding to a photon energy $\omega = 1.165$\,eV.  

The installation of ALPS~II began in 2019 and its first science run, exploiting a heterodyne detection 
scheme~\cite{Hallal:2020ibe}, 
started in May 2023. This first run does not 
include the PC, but still reaches a sensitivity in the EM coupling $g_{a\gamma\gamma}$, resp.  $g_{aMM}$,  
by two orders of manitude better than ALPS or OSQAR, probing in a purely laboratory setup the 
ALP explanation of the spectral modulation observed in gamma rays from Galactic pulsars and 
supernova remnants~\cite{Pallathadka:2020vwu}, requiring $g_{a\gamma\gamma}\simeq 2\times 10^{-10}$\,GeV$^{-1}$ at $m_a\simeq 4$\,neV.
The full optical system will be installed and used in the second half of 2023, and a further science run with upgraded optics is planned for 2024. 
The further scheduling depends on the outcome of the first science runs, results of ongoing R\&D, resources, and the news from other
axion experiments around the globe.  
It might include a science run based on an independent, single photon detection scheme exploiting 
a transition edge sensor~\cite{Dreyling-Eschweiler:2015pja,Shah:2021wsp}. 
Further options are axion searches with optimized optics and/or extension of the mass reach, 
vacuum magnetic birefringence measurements, and a dedicated search for High-Frequency Gravitational Waves 
(HFGWs)~\cite{Ejlli:2019bqj,Ringwald:2020ist}.

\subsection{Searching for Solar Axions}

\begin{figure}[h]
\begin{minipage}{0.49\linewidth}
\centerline{\includegraphics[width=\linewidth]{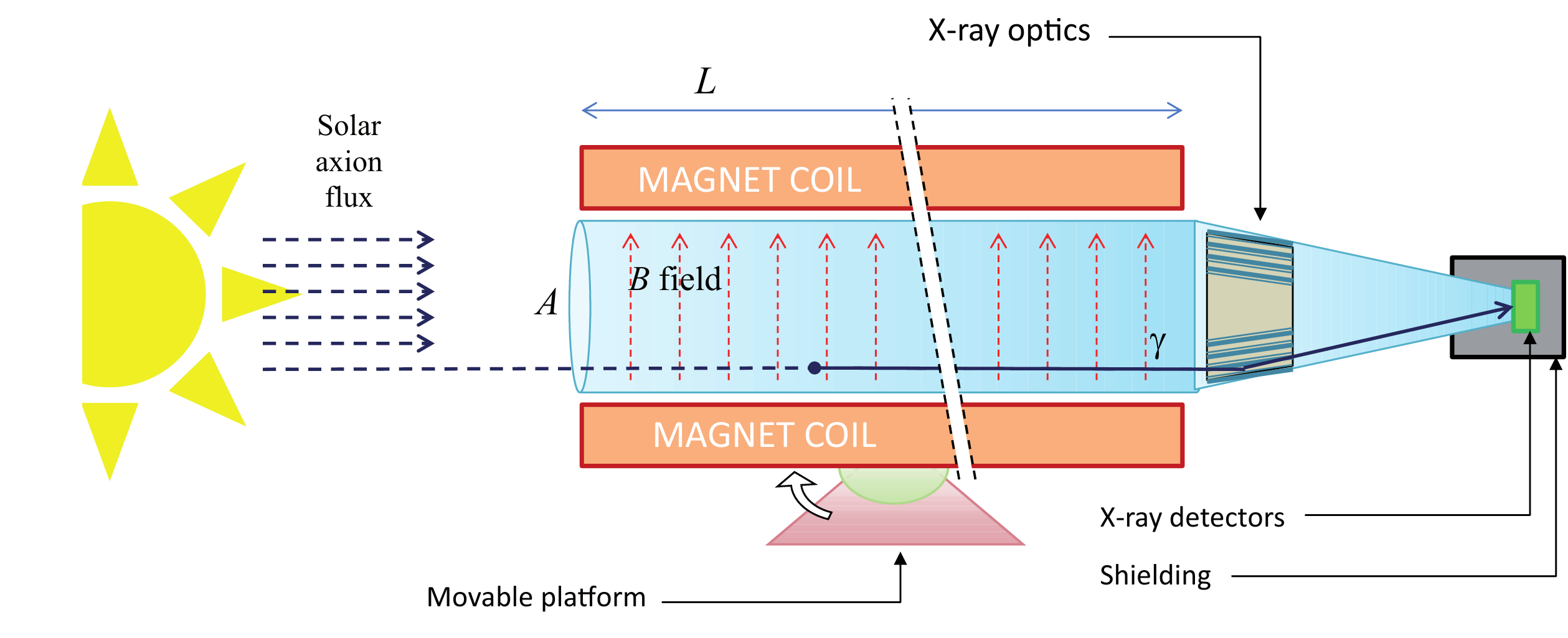}}
\end{minipage}
\hfill
\begin{minipage}{0.49\linewidth}
\centerline{\includegraphics[width=\linewidth]{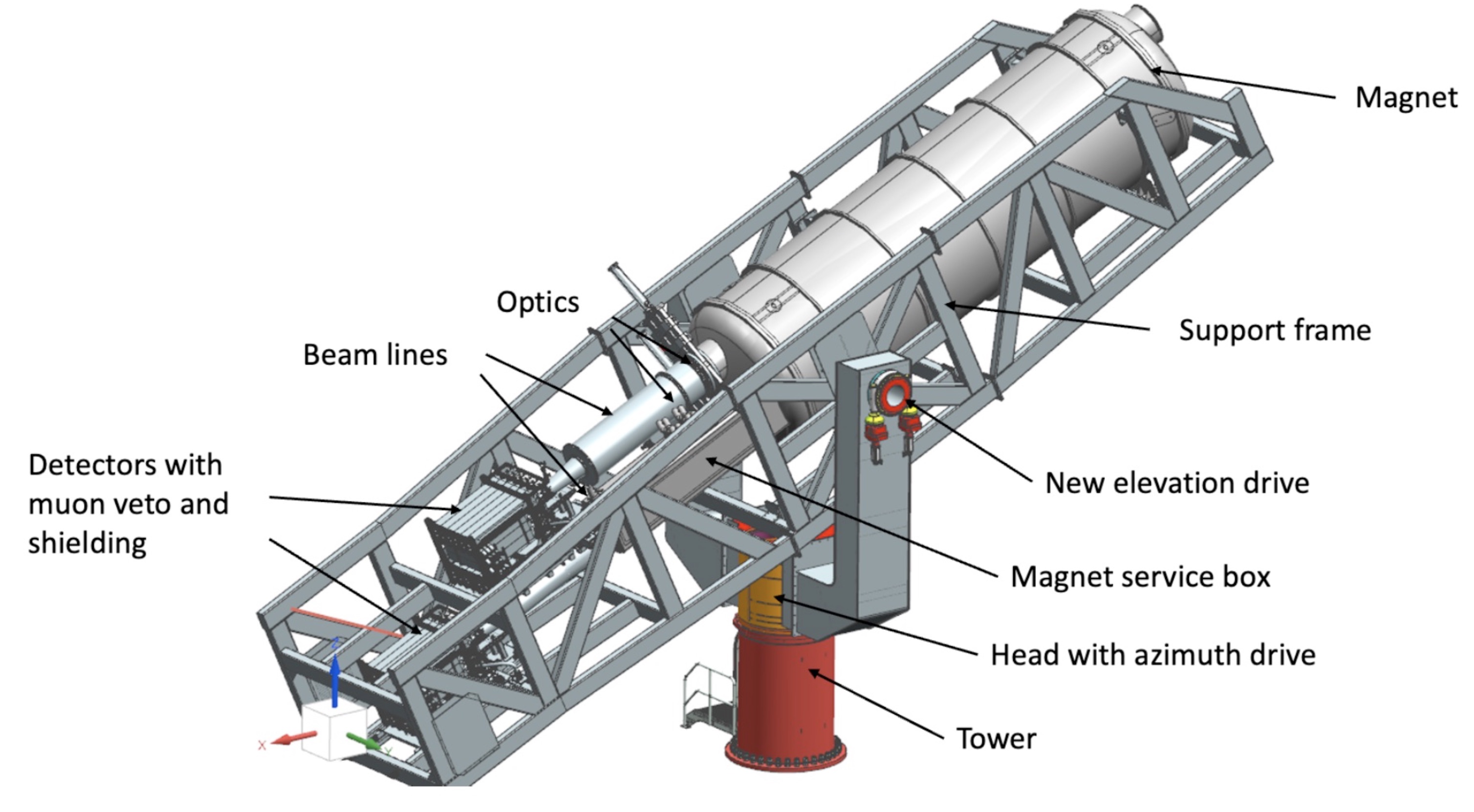}}
\end{minipage}
\caption[]{Axion helioscope concept~\cite{Armengaud:2014gea} ({\em  left}) and CAD overview of the full BabyIAXO assembly ({\em right}).}
\label{fig:BabyIAXO}
\end{figure}

The Primakoff effect in the solar plasma ($T\sim$\,keV) -- the production of axions through the interaction of 
X-ray photons with the Coulomb-field of nuclei -- may lead to a sizeable flux of solar axions, which can be searched for at Earth with an  
axion helioscope~\cite{Sikivie:1983ip}: a long dipole magnet pointed towards the 
sun in which solar axions are partially converted into photons which can be focused then on an  X-ray detector 
(see Fig.~\ref{fig:BabyIAXO} (left)).   
%
%
The CERN Axion Solar Telescope (CAST) has established an upper limit~\cite{CAST:2017uph} on $g_{a\gamma\gamma}$, resp. $g_{aMM}$, which, for 
$m_a\lesssim 10$\,meV, coincides with the HB limit in Fig.~\ref{fig:electromagnetic_couplings_vs_mass}. 
Its successor will be BabyIAXO~\cite{IAXO:2020wwp}, a prototype for the International Axion Observatory~\cite{Armengaud:2014gea} 
(IAXO). 
BabyIAXO will be based on a 10\, m long superconducting magnet ($\sim 2$\,T) with two bores, each with a diameter of 70\,cm. 
The two detection lines will feature both X-ray optics and an ultra-low background X-ray detector   
(see Fig.~\ref{fig:BabyIAXO} (right)).
BabyIAXO is designed to exceed the sensitivity of CAST on $g_{a\gamma\gamma}$, resp.  $g_{aMM}$, by a factor of around four in 
the same axion mass range (see Fig.~\ref{fig:electromagnetic_couplings_vs_mass}). 
At $m_a\lesssim 0.1$\,meV, its projected sensitivity is slightly better than the one of ALPS II, but in contrast to the latter it will probe also the meV mass ``Vanilla Axion'', and not only the ``Monopole-philic Axion'' (see Fig.~\ref{fig:electromagnetic_couplings_vs_mass}).  
Furthermore, BabyIAXO can  probe also the axion-electron  
and axion-nucleon 
couplings~\cite{Jaeckel:2018mbn,DiLuzio:2021qct}.

After the approval of BabyIAXO to be hosted at DESY, the collaboration has already started to prepare its construction. The first data taking exploiting the full BabyIAXO experiment is foreseen for 2028, although an earlier commissioning of all the subsystems except the magnet is expected for a search for solar hidden photons.
At a later point in time and by the accommodation of additional equipment, like cavities and microwave antennas, 
the BabyIAXO magnet could be used to search for axion dark matter axions~\cite{IAXO:2019mpb} and 
HFGWs~\cite{Ejlli:2019bqj,Ringwald:2020ist}.

\subsection{Searching for Dark Matter Axions}

The other axion experiments in Hamburg are axion haloscopes~\cite{Sikivie:1983ip}: they rely on the assumption that 
the dark matter halo of the Milky Way is comprised entirely by axions and search for the latter via their interaction 
with electromagnetism. The velocity dispersion of the dark-matter axions is then given by the galactic virial 
velocity, $v_a\sim 10^{-3}$, implying a macroscopic de Broglie wave length, 
$\lambda_{\rm dB} = 2\pi/(m_a v_a)\simeq {\rm km}\, (\mu{\rm eV}/m_a)(10^{-3}/v_a)$. Correspondingly, axion dark matter 
behaves as an approximately spatially homogeneous and monochromatic classical oscillating field,  
$a(t) \simeq \sqrt{2\rho_{\rm DM}} \cos (m_a t)/m_a$.\\

\noindent 
{\em WISPLC}  

\begin{figure}[h]
\vspace{-16ex}
\centerline{\includegraphics[angle=270,width=\linewidth]{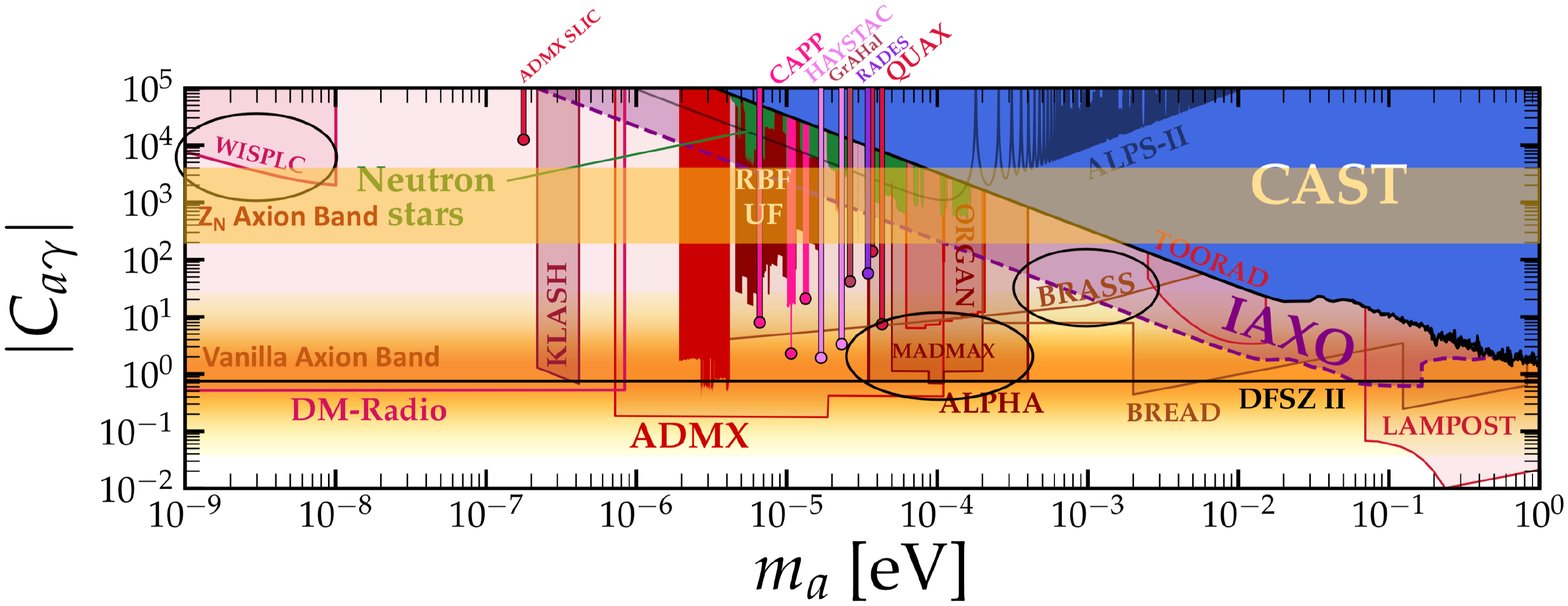}}
\vspace{-16ex}
\caption[]{Axion-photon coupling  
$C_{a\gamma}\equiv (2\pi/\alpha) (\sqrt{z}/(1+z))\,(m_\pi \, f_\pi/m_a ) \, g_{a\gamma\gamma}$ 
versus axion mass~\cite{AxionLimits}. 
The yellow region labeled ``Vanilla Axion Band" corresponds to predictions of variants of the KSVZ axion as in Fig.~\ref{fig:electromagnetic_couplings_vs_mass}, the one labeled 
``$\mathbb{Z}_N$ Axion Band" gives the dark-matter favored region of the  $\mathbb{Z}_{\mathcal N}$ axion~\cite{DiLuzio:2021gos}. 
The red filled solid regions are excluded by axion haloscopes, the regions limited by red lines are projected sensitivities of axion haloscopes. The ellipses emphasize the axion haloscopes in Hamburg.}
\label{fig:axion_discovery_potential_hamburg}
\end{figure}

A powerful approach to search for very light, $m_a\ll \mu$eV, dark matter axions~\cite{Sikivie:2013laa} can be based on the fact that, 
in the presence of a solenoidal magnetic field $\vec B$, the axion dark matter field induces an oscillating effective displacement 
current, ${\vec j}_a = - g_{a\gamma\gamma} \vec B \dot a$, which in turn generates a toroidal oscillating magnetic field 
${\vec B}_a$,  such that ${\vec\nabla}\times {\vec B}_a = {\vec j}_a$. 
%
\begin{figure}[h]
\begin{center}
\begin{minipage}{0.3\linewidth}
\centerline{\includegraphics[width=\linewidth]{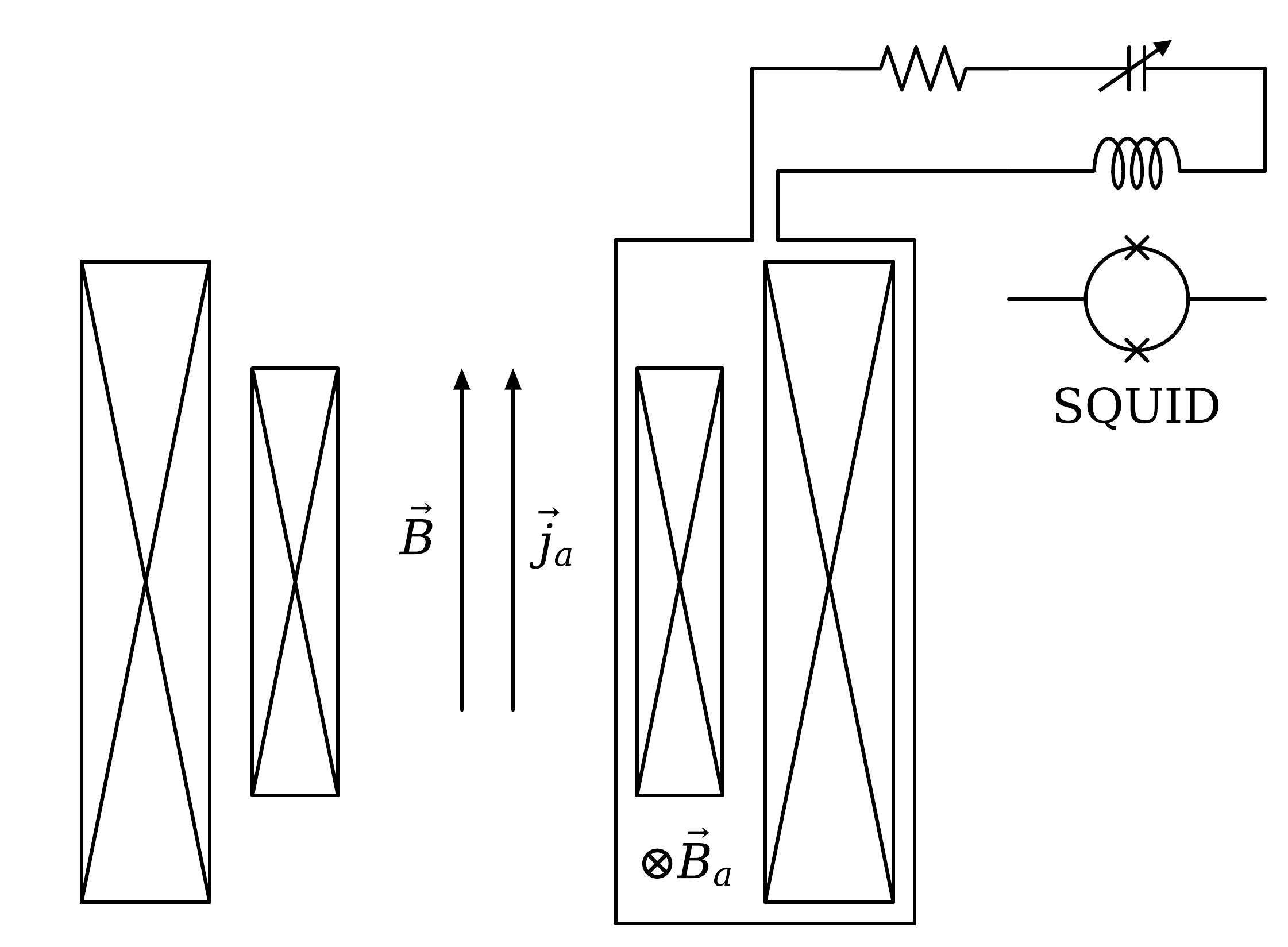}}
\end{minipage}
\hspace{12ex}
\begin{minipage}{0.18\linewidth}
\centerline{\includegraphics[width=\linewidth]{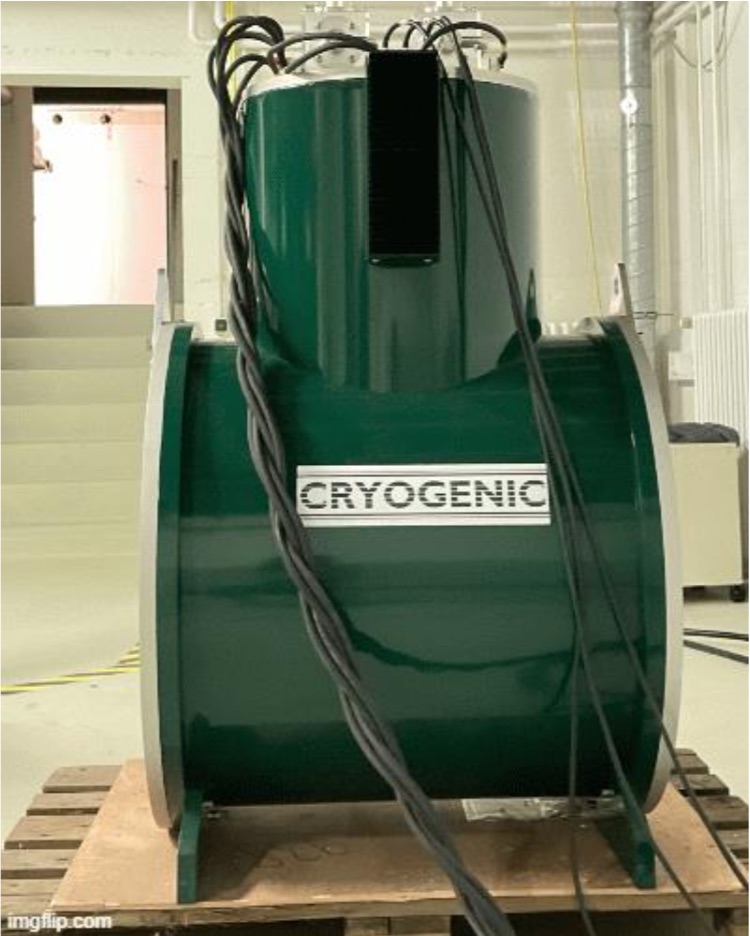}}
\end{minipage}
\end{center}
\caption[]{LC circuit haloscope concept~\cite{Zhang:2021bpa} ({\em  left}) and picture of the solenoidal magnet for WISPLC ({\em right}).}
\label{fig:WISPLC}
\end{figure}
%
The induced EM field can be turned into 
an AC in a pickup loop, resonantly amplified in a tunable LC circuit, and finally detected via a SQUID, see Fig.~\ref{fig:WISPLC} (left). 
A pilot experiment of this type, ADMX SLIC~\cite{Crisosto:2019fcj}, has obtained an upper limit on $|g_{a\gamma\gamma}|$ around $10^{-12}$\,GeV$^{-1}$,  
in a narrow mass range around $0.18\,\mu$eV, see Fig.~\ref{fig:axion_discovery_potential_hamburg}. 
%
The Weakly Interacting Slim Particle detection with LC circuit (WISPLC) experiment~\cite{Zhang:2021bpa} at the University of Hamburg 
will exploit two concentric solenoids wrapped in superconducting wire, that can produce a maximum magnetic field of 14 T at the center 
of a warm bore with a diameter of 125 mm and a length of 755 mm, see Fig.~\ref{fig:WISPLC} (right). 
The whole WISPLC experiment is fully funded and currently in the construction phase.
Its projected sensitivity is around $|g_{a\gamma\gamma}|\approx 10^{-15}~\mathrm{GeV}^{-1}$, in the axion mass range between $10^{-11}$\,eV and $10^{-8}$\,eV, still far above the expectations for ``Vanilla Axions'', 
but nearly reaching the values preferred by the trapped misalignment axion dark matter scenario of the $\mathbb{Z}_{\mathcal N}$ axion model~\cite{DiLuzio:2021gos}, see Fig.~\ref{fig:axion_discovery_potential_hamburg}.
WISPLC is also sensitive to HFGWs~\cite{Domcke:2023bat,Domcke:2023qle}, but it is 
not sensitive to the $g_{aMM}$ coupling of the monopole-philic axion model~\cite{Li:2022oel}.
However, with a simple change from a toroidal pick-up loop to a solenoidal one~\cite{Li:2022oel} 
it can measure the CP violating $g_{aEM}$
 coupling~\cite{Sokolov:2022fvs,Sokolov:2023pos} which arises if  
the exotic quark $\mathcal Q$ in the triangle loop in Fig.~\ref{fig:electromagnetic_coupling_axion}  
features both a magnetic as well as an electric charge, that is if it is not only a monopole, but rather a dyon.\\

\noindent
{\em BRASS}

The dish antenna axion haloscope concept has been proposed~\cite{Horns:2012jf} as a new broadband search method for dark matter axions with higher masses, $m_a \gg \mu$eV. It is based on the fact that the oscillating axion DM, in a background magnetic field $\vec B$, carries an oscillating electric field component parallel to the latter, ${\vec E}_a(t) = - g_{a\gamma\gamma} \vec B a(t)$. 
\begin{figure}[h]
\begin{center}
\begin{minipage}{0.3\linewidth}
\centerline{\includegraphics[width=\linewidth]{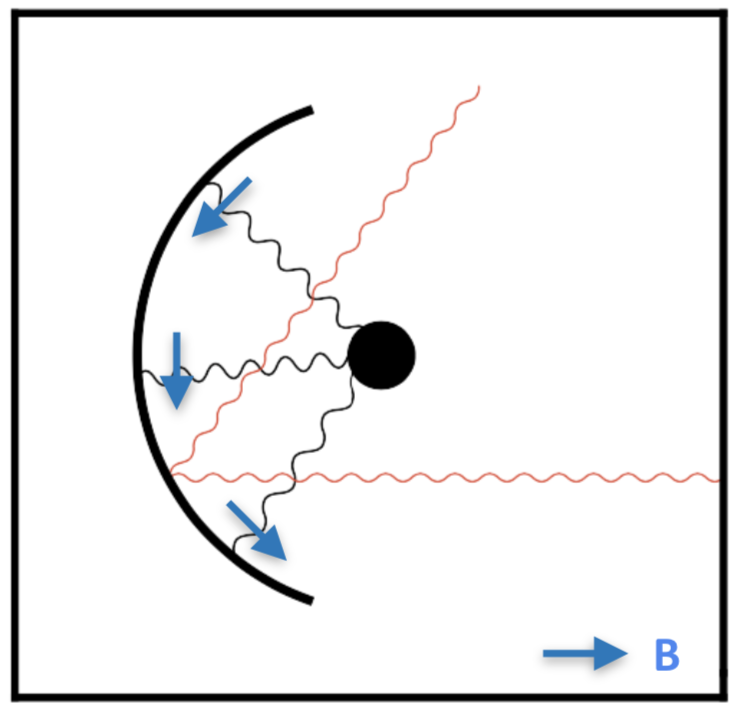}}
\end{minipage}
\hspace{12ex}
\begin{minipage}{0.45\linewidth}
\centerline{\includegraphics[width=\linewidth]{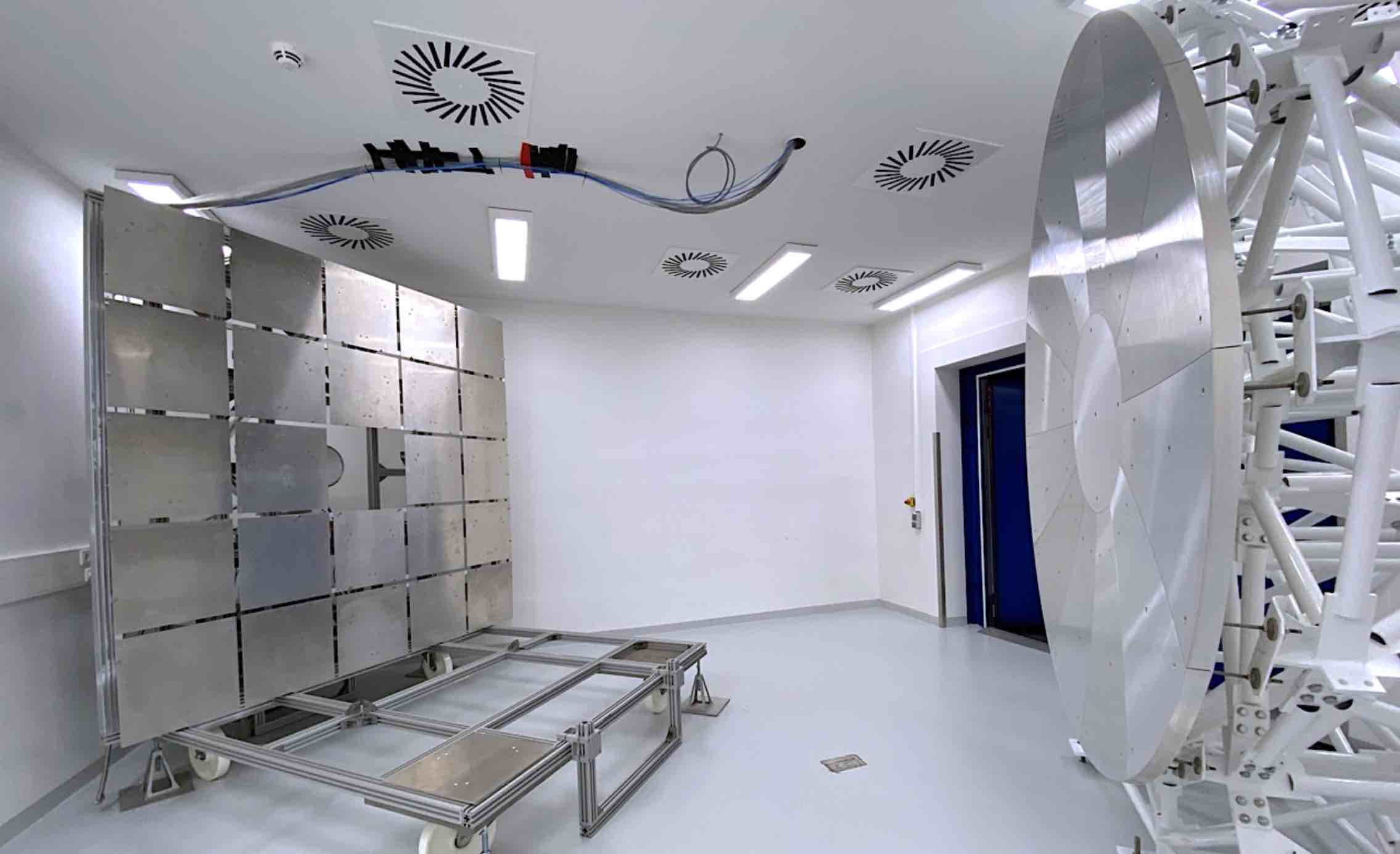}}
\end{minipage}
\end{center}
\caption[]{Dish antenna axion haloscope concept~\cite{Horns:2012jf}  ({\em  left}) and picture of BRASS ({\em right}).}
\label{fig:BRASS}
\end{figure}
This leads to the effect that a metallic mirror placed in a magnet field pointing parallel to the mirror surface will emit a
nearly monochromatic 
EM wave perpendicular to the mirror surface with a frequency $\nu = m_a/(2\pi)$ and a cycle-averaged 
power per unit area, 
${{\mathcal P}_\gamma}/{\mathcal A}=|\vec{E}_a|^2/2=2.2\times 10^{-27}\, {\rm W/m^2} |C_{a\gamma}|^2 (|\vec B|/(10\,{\rm T}))^2$,
that is amenable to detection, see Fig.~\ref{fig:BRASS} (left).
The Broadband Radiometric Axion SearcheS (BRASS) experiment~\cite{Bajjali:2023uis} at the University of Hamburg exploits $24 \times 0.25$\,m$^2$ flat permanent magnetic (0.8\,T) conversion panels, a parabolic mirror to collect the signal and focus it onto a broadband (12-18\,GHz) 
receiver with a cryogenic frontend. This pilot dish antenna axion haloscope is expected to reach, in $\sim 10^3$ hours measurement time, a sensitivity of $|C_{a\gamma}|\approx 10^3$,
 in the mass range $\sim (50-70)\,\mu$eV, probing the $\mathbb{Z}_{\mathcal N}$ axion parameter space, see Fig.~\ref{fig:axion_discovery_potential_hamburg}. In future upgrades, BRASS is planned to touch the ``Vanilla Axion'' parameter space, 
$|C_{a\gamma}|\lesssim 10$, see
Fig.~\ref{fig:axion_discovery_potential_hamburg}.\\

\noindent 
{\em MADMAX}

A dielectric haloscope~\cite{Caldwell:2016dcw} is essentially a boosted dish antenna axion haloscope: 
it consists of a mirror and a series of parallel, partially transparent dielectric disks in front of it, all within a magnetic field parallel to the surfaces, 
and a receiver in the field-free region, as 
\begin{figure}[h]
\vspace{-8ex}
\begin{minipage}{0.37\linewidth}
\centerline{\includegraphics[width=\linewidth]{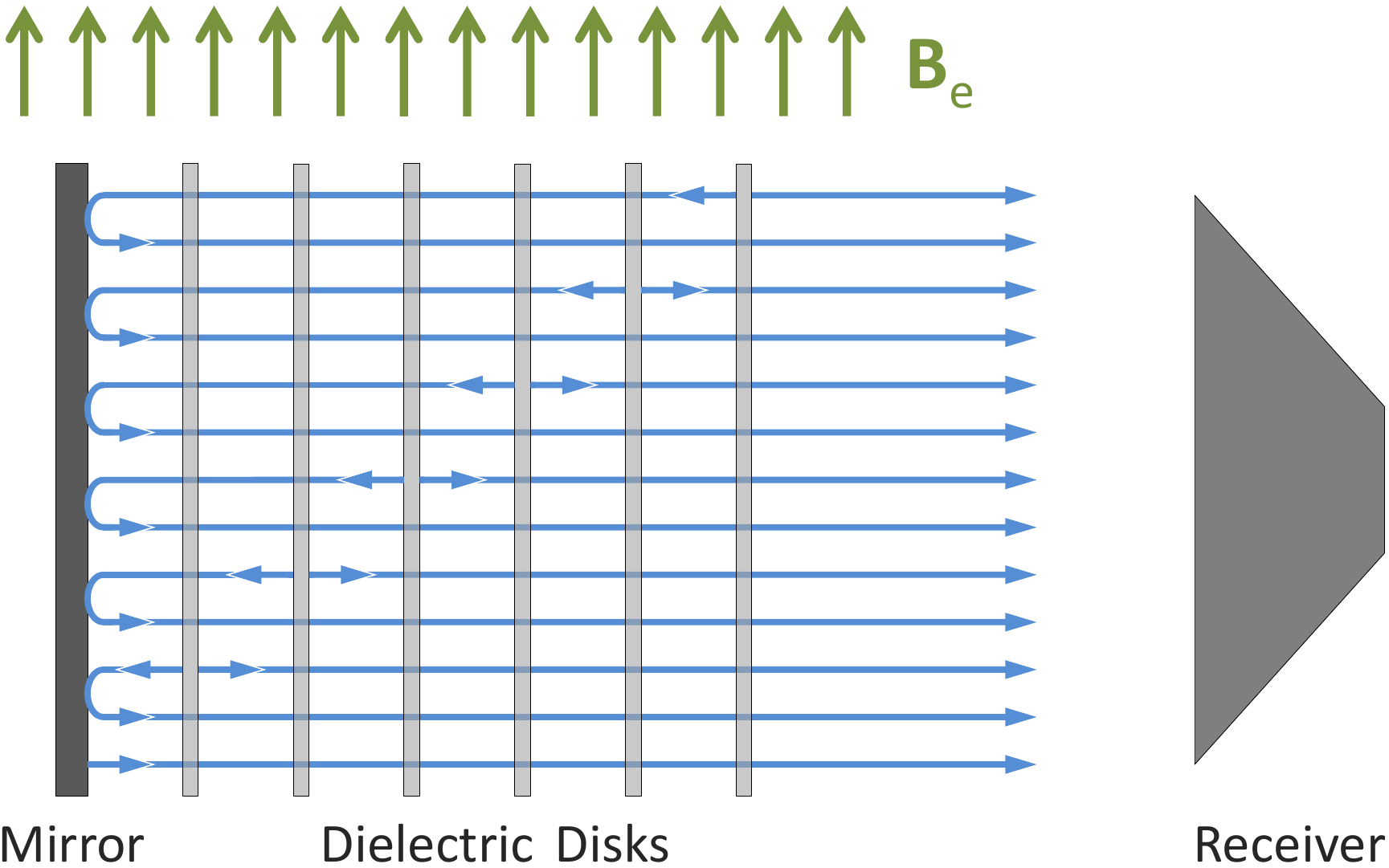}}
\end{minipage}
\hfill
\begin{minipage}{0.63\linewidth}
\centerline{\includegraphics[angle=270,width=\linewidth]{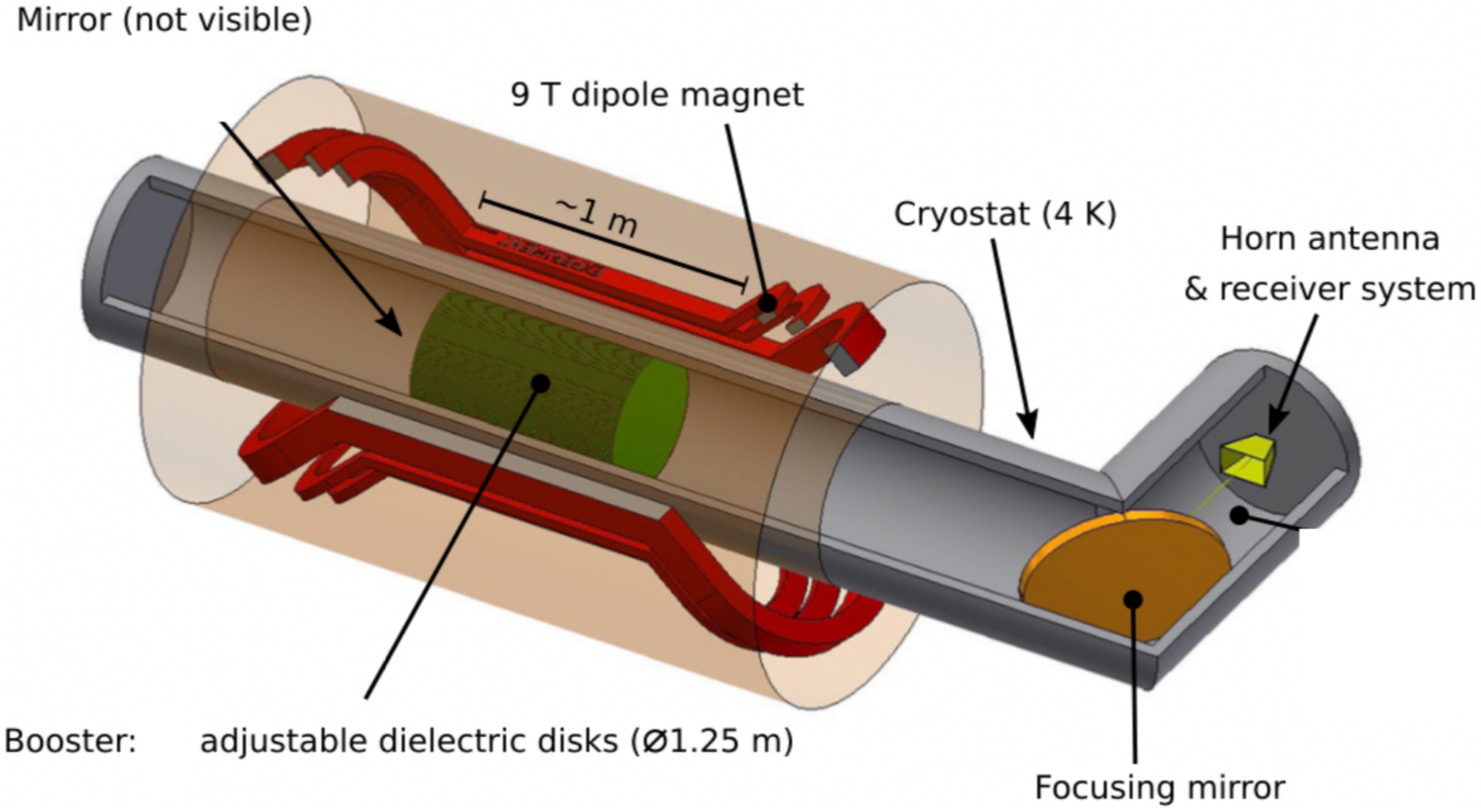}}
\end{minipage}
\vspace{-8ex}
\caption[]{Boosted dish antenna aka as 
dielectric haloscope concept~\cite{Caldwell:2016dcw} ({\em  left}) and sketch of MADMAX ({\em right}).}
\label{fig:madmax}
\end{figure}
shown in  Fig.~\ref{fig:madmax}~(left). Each disk acts as a flat dish antenna. The waves emitted by each disk are reflected by and transmitted through the other disks before exiting. For suitable disk separations, these waves add coherently to enhance the emitted power.    
This allows scans over a band of $m_a$ without needing to use disks with different thicknesses for each measurement.

The MAgnetized Disks and Mirror Axion eXperiment (MADMAX) collaboration~\cite{MADMAX:2019pub} aims at building an 
adjustable multiple-disk system (booster) of $\mathcal A \sim  1$\,m$^2$ inside a $\sim 9$\,T dipole magnet. 
With an expected power boost factor of $\beta^2(\nu )\sim 10^4$ and equipped with a quantum-limited receiver, it is projected 
to scan the $(40-400)\,\mu$eV mass range with DFSZ sensitivity, as shown in Fig.~\ref{fig:axion_discovery_potential_hamburg}. 
MADMAX is planned to be located at DESY in HERA's North Hall, near  ALPS II. 

Currently, it is in the R\&D phase. 
At CERN, the collaboration is using the MORPURGO magnet with a dipole field of up to 1.6\,T for a prototype experiment,
mainly to check the performance of the booster.
This allows for physically interesting and competitive axion dark matter searches, reaching $\mathbb{Z}_{\mathcal N}$ axion sensitivity around 
$78.5\,\mu$eV. 
The schedule of the full MADMAX experiment is mainly determined by the  funding decision for the large dipole magnet. Data taking in Hamburg may start in 2030, but already earlier a new prototype magnet could allow for axion dark matter 
and HFGW~\cite{Ringwald:2020ist} searches in mass and frequency ranges which are so far largely unexplored.

\section*{Acknowledgments}
Special thanks to E.~Garutti, I.~Irastorza, L.H.~Nguyen, J.~Schaffran, C.~Schwemmbauer and A.~Sokolov for valuable comments on the draft.
This work has been partially funded by the Deutsche Forschungsgemeinschaft (DFG, German Research Foundation) 
under Germany’s Excellence Strategy - EXC 2121 Quantum Universe - 390833306  and under 
- 491245950.

\end{document}